\documentclass[a4paper,11pt]{article}

\usepackage{jheppub}
\usepackage{graphicx}
\usepackage{bm}
\usepackage{hyperref}

\newcommand{\Tr}{{\rm Tr}}
\newcommand{\myslash}[1]{#1\!\!\!/}

\title{$T_{cc}^+$ and $X(3872)$ with the complex scaling method and $DD(\bar{D})\pi$ three-body effect}

\author{Zi-Yang Lin}
\author{Jian-Bo Cheng}
\author{Shi-Lin Zhu}

\affiliation{
  School of Physics and Center of High Energy Physics, Peking University 10087, China
}%

\emailAdd{lzy\_15@pku.edu.cn}
\emailAdd{jbcheng@pku.edu.cn}
\emailAdd{zhusl@pku.edu.cn}

\abstract{We use the leading order (LO) contact interactions and OPE potentials to investigate the newly observed double-charm state $T_{cc}^+$. The $DD\pi$ three-body effect is important in this system since the intermediate states can go on shell. We keep the dependence of the pion propagators on the center-of-mass energy, which results in a unitary cut of the OPE potential at the $DD\pi$ three-body threshold. By solving the complex scaled Schr\"odinger equation, we find a pole corresponding to the $T_{cc}^+$ on the physical Riemann sheet. Its width is around 80 keV and nearly independent of the choice of the cutoff. Assuming the $D\bar{D}\pi$ and $D\bar{D}^*$ channels as the main decay channels, we apply the similar calculations to the $X(3872)$, and find its width is even smaller. Besides, the isospin breaking effect is significant for the $X(3872)$ while its impact on the $T_{cc}^+$ is relatively small.}

\begin{document}
\maketitle
\flushbottom

\section{Introduction}
Recently, the LHCb Collaboration observed a double-charm exotic hadron with $J^P=1^+$ named as $T_{cc}^+$ in the $D^0D^0\pi^+$ mass spectrum, and its mass and width are \cite{LHCb:2021vvq}
\begin{eqnarray}
  & \delta m_{\text{BW}}=-273\pm 61\text{ keV}, \quad &\Gamma_{\text{BW}}=410\pm 165\text{ keV},  
\end{eqnarray}
where $\delta m_\text{BW}$ is the $T_{cc}^+$ mass shift with respect to the $D^0D^{*+}$ threshold and $\Gamma$ denotes its width. This result is extracted from a relativistic P-wave two-body Breit-Wigner parameterization and is only a rough description of the state. Since the state is rather close to the $D^0D^{*+}$ threshold, a further study is carried out using a unitarized Breit-Wigner parameterization, which gives \cite{LHCb:2021auc}
\begin{eqnarray}
  & \delta m_{\text{U}}=-359\pm 40\text{ keV}, \quad &\Gamma_{\text{U}}=47.8\pm 1.9\text{ keV},\nonumber\\  
  & \delta m_{\text{pole}}=-360\pm 40\text{ keV}, \quad &\Gamma_{\text{pole}}=48\pm 2\text{ keV}.
\end{eqnarray}
The pole position is also derived from the unitarized Breit-Wigner parameterization, which is directly related to our theoretical calculations.

Another important feature of the $T_{cc}^+$ is the absence of signals in the $D^+D^{*+}$ and $D^+D^0\pi^+$ mass spectra. It implies that the $T_{cc}^+$ is an isoscalar rather than an isovector, which is similar to the charmonium-like exotic state $X(3872)$ first observed in 2003 \cite{Belle:2003nnu}, with the spin-parity quantum number $J^{PC}=1^{++}$ and isospin $I=0$. Since the $X(3872)$ lies too close to the $D^0\bar{D}^{*0}$ threshold, the LHCb Collaboration made a more precise investigation of its lineshape in 2020 \cite{LHCb:2020xds}. The generic Breit-Wigner mass and width are
\begin{eqnarray}
  &m_{\chi_{c1}(3872)}=3871.695 \pm 0.067 \pm 0.068 \pm 0.010\text{ MeV},\nonumber\\
  &\Gamma_{\text{BW}}=1.39\pm 0.24 \pm 0.10\text{ MeV}.
\end{eqnarray}
Due to the proximity to the threshold, the fit with the Flatt\'e lineshape is also carried out, and the peak position and the full width of the half maximum is
\begin{eqnarray}
  &\text{mode}=3871.69^{+0.00+0.05}_{-0.04-0.13}\text{ MeV},\nonumber\\
  &\text{FWHM}=0.22^{+0.07+0.11}_{-0.06-0.13}\text{ MeV}.
\end{eqnarray} 
The pole search is also done, and a pole on the first (physical) Riemann sheet with respect to the $D^0\bar{D}^{*0}$ threshold is preferred. At the best estimate of the Flatt\'e parameters, the pole position is found to be
\begin{eqnarray}
  E=0.06-0.13i\text{ keV},
\end{eqnarray}
where the imaginary part of the pole position corresponds to the opposite of half the width. Notice that the generic Breit-Wigner parameterization may overestimate the widths of the states close to thresholds, and the peak positions do not correspond to the physical pole positions. Here we always focus on the pole positions.

A variety of works have been put into effort to study the double- and hidden-charm tetraquarks. There are several theoretical interpretations for the double-charm tetraquarks: the molecular state picture \cite{Manohar:1992nd,Janc:2004qn,Ohkoda:2012hv,Chen:2021vhg,Chen:2021cfl,Chen:2021spf,Dong:2021bvy,Deng:2021gnb,Feijoo:2021ppq,Dai:2021vgf,Deng:2022cld,Albaladejo:2021vln}, the compact tetraquark picture \cite{Ballot:1983iv,Zouzou:1986qh,Yang:2009zzp,Berezhnoy:2018bde,Yang:2019itm,Tan:2020ldi,Guo:2021yws,Meng:2021yjr}, the QCD sum rule \cite{Du:2012wp} and the lattice QCD simulation \cite{Cheung:2017tnt,Junnarkar:2018twb}. As for the hidden-charm $X(3872)$, see Refs.~\cite{Chen:2016qju,Esposito:2016noz,Guo:2017jvc,Liu:2019zoy,Brambilla:2019esw,Chen:2022asf} for a detailed review. 

In this work, we regard the $T_{cc}^+$ ($X(3872)$) as a loosely bound $DD^* (D\bar{D}^*)$ molecular state. The chiral effective field theory (ChEFT) serves as a helpful tool to study this system since the momenta of the charmed mesons are small. Based on Weinberg’s power counting \cite{Weinberg:1990rz,Weinberg:1991um}, the calculations can be organized by the powers of the small external momenta and pion mass. In the ChEFT, the one-pion-exchange (OPE) potential provides the long-range attraction. Together with the intermediate-range interactions from the two-pion-exchange (TPE) and short-range interactions from the contact terms, the ChEFT provides a successful description of the nucleon systems \cite{Bernard:1995dp,Machleidt:2011zz}. Similar to the heavy baryon chiral effective field theory used in the nucleon systems \cite{Gasser:1987rb}, the heavy meson chiral effective field theory (HMChEFT) is performed to deal with the charmed mesons \cite{Wise:1992hn}. For a review of the ChEFT for heavy hadronic molecules, see Ref.~\cite{Meng:2022ozq}.

However, the $DD^*$ system is somehow extraordinary because the mass splitting between the pseudoscalar meson $D$ and the vector meson $D^*$ is slightly larger than the pion mass. Therefore, the exchanged pion can go on shell, which makes the instantaneous approximation (transferred energy $p^0=0$) inappropriate in the $DD^*$ system. If we keep the pion energy $p^0\approx m_{D^*}-m_D$ instead, a pole singularity shows up in the OPE potential, which gives rise to a non-vanishing imaginary part. Some of the previous works \cite{Li:2012cs} kept the non-vanishing $p^0$ but dropped the imaginary part of the potential through a principal integral when performing the Fourier transformation to the coordinate space. Here we keep the imaginary part and revise the OPE potential to consider the $DD\pi$ three-body threshold effect. As we will see, the Hamiltonian is no longer Hermitian, which leads to the complex energy eigenvalues.

Following Ref.~\cite{Baru:2011rs}, a few works discussed the three-body dynamics from the point of view of the coupled-channel approach in quantum mechanics \cite{Schmidt:2018vvl,Du:2021zzh}. A non-relativistic propagator $(E-H_0+i\epsilon)^{-1}$ is used for the intermediate three-body states, where the $E$ denotes the center-of-mass energy, and the $H_0$ denotes the total kinetic energy of the intermediate $DD\pi$ systems. Thus the OPE potentials become dependent on the center-of-mass energy. In this way, a unitary cut at the $DD\pi$ threshold is introduced. They found that the static pion approximation ($p^0=0$) overestimates the width.

Different from previous works, we shall keep the relativistic form of the pion propagators in the quantum field theory (Feynman prescription), with a proper selection of the 0-th component of the exchanged pion momentum. The unitary cut at the $DD\pi$ threshold is introduced through the energy dependence. Notice that the $DD\pi$ final state is the only allowed strong decay channel of the $T_{cc}^+$, which has been included in the OPE potential, or the $DD^*\pi$ vertex. In this work, we will use the potential involving the three-body effects and the complex scaling method (CSM) to obtain the binding energies and widths at the same time. 

The complex scaling method is a useful similarity transformation of the Schr\"odinger equation \cite{Aguilar:1971ve,Balslev:1971vb}, which allows us to obtain the bound states and resonances at the same time by directly solving the complex scaled Schr\"odinger equation. For a review of applications of the CSM in the nucleus systems, see Ref.~\cite{Myo:2014ypa}. As a tool of the analytical continuation, the CSM can also help us deal with the singularity of the potential.

This paper is organized as follows. In Sec.~\ref{sec:formalism}, we introduce the chiral Lagrangians, the complex scaling method and the classification of the poles related to the CSM. We explain why we revise the OPE potential and how to handle the analytical continuation correctly. In Sec.~\ref{sec:Veff}, we present the effective potentials explicitly, with the isospin breaking effect. In Sec.~\ref{sec:numeric}, we show the numerical results of the $T_{cc}^+$ and $X(3872)$. In Sec.~\ref{sec:TPE}, we investigate the influence of the TPE potential to the width. In Sec.~\ref{sec:sum}, we make a summary.

\section{Formalism\label{sec:formalism}}
\subsection{Chiral Lagrangian}

In the HMChEFT, the Lagrangians, together with the scattering amplitudes, can be organized in the powers of the small external momenta $q$ over a large energy scale $\Lambda_\chi\sim 1\text{ GeV}$. The $\Lambda_\chi$ represents the chiral breaking scale. Up to the $\mathcal{O}(p^2)$ amplitudes, we need only the leading-order (LO) Lagrangian for the $H\phi$ interaction
\begin{eqnarray}
  \mathcal{L}_{H\phi}^{(1)}&=-\langle(iv\cdot\partial H)\bar{H}\rangle+\langle Hv\cdot\Gamma\bar{H}\rangle\nonumber\\
  &+g\langle H\myslash{u}\gamma_5\bar{H}\rangle-\frac{1}{8}\delta\langle H\sigma^{\mu\nu}\bar{H}\sigma_{\mu\nu}\rangle,\label{eq:Hphi1} 
\end{eqnarray}
where $\phi$ denotes the Goldstone bosons, and $H$ denotes the heavy meson doublet under the heavy quark spin symmetry (HQSS) \cite{Wise:1992hn}, which is defined as
\begin{eqnarray}
  &&H=\frac{1+\myslash{v}}{2}(P_\mu^*\gamma^\mu+iP\gamma_5),\nonumber\\
    &&\bar{H}=\gamma^0H^\dagger\gamma^0=(P_\mu^{*\dagger}\gamma^\mu+iP^\dagger\gamma_5)\frac{1+\myslash{v}}{2},\\
    &&P=(D^0,D^+),\quad P_\mu^*=(D^{*0},D^{*+}).\nonumber
\end{eqnarray}
The last term in Eq.~(\ref{eq:Hphi1}) introduces the mass splitting between heavy pseudoscalar and vector mesons, and $v=(1,\vec{0})$ is the $4$-velocity of the heavy mesons. The chiral connection $\Gamma_\mu$ and the axial current $u_\mu$ contain an even and odd number of the Goldstone bosons, respectively, which read

\begin{eqnarray}
  &&\Gamma_\mu=\frac{i}{2}[\xi^\dagger,\partial_\mu\xi]=-\frac{1}{4f_\pi^2}\epsilon^{abc}\tau^c(\phi^a\partial_\mu \phi^b)+\cdots,\nonumber\\
    &&u_\mu=\frac{i}{2}\{\xi^\dagger,\partial_\mu\xi\}=-\frac{1}{2f_\pi}\tau^a\partial_\mu \phi^a+\cdots\nonumber,\\
    &&\xi=\exp(i\phi/2f_\pi),\\
    &&\phi=\phi^a\tau^a=\sqrt{2}\begin{pmatrix}
        \frac{\pi^0}{\sqrt{2}}&\pi^+\\
        \pi^-&-\frac{\pi^0}{\sqrt{2}}
    \end{pmatrix},\nonumber
\end{eqnarray}
where $\lambda^a$ denotes the Pauli matrices, and $f_\pi$ represents the decay constant of the Goldstone bosons.

The next-to-leading-order (NLO) $H\phi$ Lagrangian $\mathcal{L}_{H\phi}^{(2)}$ contains at least two light mesons, which only appears in the TPE diagrams and thus does not contribute to the $\mathcal{O}(p^2)$ amplitudes.

To mimic the short-range interactions between the heavy mesons, we need the contact Lagrangian 
\begin{eqnarray}
  &\mathcal{L}^{(0)}_{4H}&=D_a\Tr[H\gamma_\mu\bar{H}]\Tr[H\gamma^\mu\bar{H}]\nonumber\\
  &&+D_b\Tr[H\gamma_\mu\gamma_5\bar{H}]\Tr[H\gamma^\mu\gamma_5\bar{H}]\nonumber\\
  &&+E_a\Tr[H\gamma_\mu\tau^a\bar{H}]\Tr[H\gamma^\mu\tau^a\bar{H}]\nonumber\\
  &&+E_b\Tr[H\gamma_\mu\gamma_5\tau^a\bar{H}]\Tr[H\gamma^\mu\gamma_5\tau^a\bar{H}].\label{eq:L4H1}
\end{eqnarray}

For the $D\bar{D}^*$ system, the Lagrangian is constructed as follows,
\begin{eqnarray}
  &\mathcal{L}^{(0)}_{4H}&=2\tilde{D}_a\Tr[\bar{\tilde{H}}\gamma_\mu\tilde{H}]\Tr[H\gamma^\mu\bar{H}]\nonumber\\
        &&+2\tilde{D}_b\Tr[\bar{\tilde{H}}\gamma_\mu\gamma_5\tilde{H}]\Tr[H\gamma^\mu\gamma_5\bar{H}]\nonumber\\
        &&+2\tilde{E}_a\Tr[\bar{\tilde{H}}\gamma_\mu\tau^a\tilde{H}]\Tr[H\gamma^\mu\tau^a\bar{H}]\nonumber\\
        &&+2\tilde{E}_b\Tr[\bar{\tilde{H}}\gamma_\mu\gamma_5\tau^a\tilde{H}]\Tr[H\gamma^\mu\gamma_5\tau^a\bar{H}],\label{eq:L4H2}
\end{eqnarray}
\begin{eqnarray}
  \mathcal{L}_{\tilde{H}\phi}^{(1)}&=-\langle(iv\cdot\partial \bar{\tilde{H}})\tilde{H}\rangle+\langle \bar{\tilde{H}}v\cdot\Gamma\tilde{H}\rangle\nonumber\\
  &+g\langle \bar{\tilde{H}}\myslash{u}\gamma_5\tilde{H}\rangle-\frac{1}{8}\delta\langle \bar{\tilde{H}}\sigma^{\mu\nu}\tilde{H}\sigma_{\mu\nu}\rangle,
\end{eqnarray}
where $\tilde{H}$ stands for the heavy anti-meson fields

\begin{eqnarray}
    &&\tilde{H}=(\tilde{P}_\mu^*\gamma^\mu+i\tilde{P}\gamma_5)\frac{1-\myslash{v}}{2},\nonumber\\
    &&\bar{\tilde{H}}=\gamma^0H^\dagger\gamma^0=\frac{1-\myslash{v}}{2}(\tilde{P}_\mu^{*\dagger}\gamma^\mu+i\tilde{P}^\dagger\gamma_5),\\
    &&\tilde{P}=\begin{pmatrix}
        \bar{D}^0\\D^-
    \end{pmatrix}
    ,\qquad \tilde{P}_\mu^*=\begin{pmatrix}
        \bar{D}^{*0}\\D^{*-}
    \end{pmatrix}.\nonumber
\end{eqnarray}
Note that the LECs in Eq.~(\ref{eq:L4H1}) and Eq.~(\ref{eq:L4H2}) are not exactly the same since we leave out the terms like $\Tr[H\bar{H}]\Tr[H\bar{H}]$. In the $DD^{*}$ system, these terms are not independent of the terms in Eq.~(\ref{eq:L4H1}). However, they do get an extra opposite sign under the charge conjugation, compared with the terms in Eq.~(\ref{eq:L4H1}). Besides, we could introduce an imaginary part to the LECs in Eq.~(\ref{eq:L4H1}) to involve the annihilation effects and contributions from the inelastic channels.

In principle, the higher order contact terms $L_{4H}^{(2)}$ should be included to cancel the divergences of the loop integrals. Although these terms do contribute to the $\mathcal{O}(p^2)$ amplitudes, we ignore their finite part due to the lack of the experimental data. In this work, we calculate the effective potentials under the heavy meson limit, and neglect all $1/M_{D^{(*)}}$ corrections.

\subsection{Complex scaled Schr\"odinger equation}
Once deriving the effective potentials, we can search for the possible bound states or resonances using the CSM. We consider the Schr\"odinger equation in momentum space
\begin{eqnarray}
  E\phi_{l}(p)=\frac{p^2}{2m}\phi_{l}(p)+\int\frac{p'^2dp'}{(2\pi)^3}V_{l,l'}(p,p')\phi_{l'}(p')\label{eq:sch1},
\end{eqnarray}
and perform the complex scaling operation
\begin{eqnarray}
    p\rightarrow pe^{-i\theta},  \qquad \tilde{\phi}_l(p)=\phi_l(pe^{-i\theta}).\label{eq:rotate}
\end{eqnarray}

Then we derive the complex scaled Schr\"odinger equation with a scaling angle $\theta$
\begin{eqnarray}
  &E\tilde{\phi}_{l}(p)&=\frac{p^2e^{-2i\theta}}{2m}\tilde{\phi}_{l}(p)\nonumber\\
  &+&\int\frac{p'^2e^{-3i\theta}dp'}{(2\pi)^3}V_{l,l'}(pe^{-i\theta},p'e^{-i\theta})\tilde{\phi}_{l'}(p'),\label{eq:sch2}
\end{eqnarray}
where $l,l'$ are the orbital angular momenta, and $p$ denotes the momentum in the center-of-mass frame.  

The eigenenergy remains unchanged after the substitution in Eq.~\ref{eq:rotate}, but with the rotation operation shown above, the poles of the $T$ matrix on the second Riemann sheet can be revealed by solving the complex scaled Schr\"odinger equation directly. This can be roughly shown using the asymptotic solution in coordinate space
\begin{eqnarray}
  \psi(r)\stackrel{r\rightarrow\infty}{\longrightarrow} f_l^+(k)e^{-ikr}+f_l^-(k)e^{ikr}.
\end{eqnarray}
With the complex scaling operation in coordinate space $r\rightarrow re^{i\theta}$, the asymptotic wavefunction becomes
\begin{eqnarray}
  \tilde{\psi}(r)\stackrel{r\rightarrow\infty}{\longrightarrow} f_l^+(k)e^{-ikre^{i\theta}}+f_l^-(k)e^{ikre^{i\theta}}.
\end{eqnarray}

 The first term vanishes at the zeros of the Jost function $f_l^+(k)$ which correspond to the poles of the $T$ matrix. We can solve the eigenstates as long as $\tilde{\psi}(r)$ converges at $r\rightarrow\infty$, i.e. $\text{Im } ke^{i\theta}>0$ or $\text{Arg }k>-\theta$.

\begin{figure}
  \centering
  \includegraphics[width=75mm]{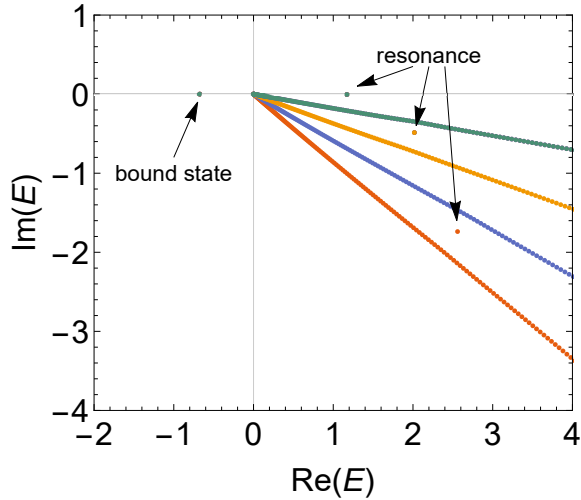}
  \caption{\label{fig:csm}A typical solution of the complex scaled Schr\"odinger equation. Eigenenergies are plotted on the complex plane. The continuum states line up due to the same arguments $\text{Arg} (E)=-2\theta$. With a Hermitian Hamiltonian, the bound states lie on the negative real axis, while the resonances lie on the fourth quadrant, and can be seen only when $|\text{Arg} (E)|<2\theta$. }
\end{figure}

A typical distribution of the poles solved by the CSM is shown in Fig.~\ref{fig:csm}. The continuum states line up while the resonances and bound states lie above and below the continuum states, respectively. In the usual cases when the Hamiltonian is Hermitian, the bound states lie accurately on the negative real axis and resonances lie on the fourth quadrant. However, the Hamiltonians are no longer Hermitian when we encounter the decay processes. As we will see in Subsec.~\ref{chap:3bodies}, the coupled-channel effects of a lower threshold may introduce a complex potential with a non-vanishing imaginary part, which moves the bound state to the third quadrant. For a precise classification of the poles, see Ref.~\cite{Badalian:1981xj}. These states are called the unstable bound states (UBS) (or quasibound state in some references).

The complex scaling method is equivalent to the Lippmann-Schwinger equation. However, it provides a more efficient way to obtain the bound states and resonances simultaneously with no need to calculate the $T$ matrix. Furthermore, we can use CSM to bypass the poles on the real axis when dealing with the OPE potentials. 

\subsection{Effects of the three-body threshold\label{chap:3bodies}}
A typical OPE potential has the form of
\begin{eqnarray}
  &V_{1\pi}&=C\frac{(\epsilon\cdot p)(\epsilon'\cdot p)}{p^2-m_\pi^2+i0+}\nonumber\\
  &&=C\frac{(\epsilon\cdot p)(\epsilon'\cdot p)}{p_0^2-\bm{p}^2-m_\pi^2+i0+},
\end{eqnarray}
where $p$ represents the transferred momentum, $C$ is a constant related to the isospin and coupling constants, and $\epsilon,\epsilon'$ represent the polarization vectors of the initial and final $D^*$ mesons, respectively.

The instantaneous approximation, namely taking $p_0=0$, will lead to the usual OPE potential without singularity. Nonetheless, it is not appropriate here since $p_0\sim \delta=m_{D^*}-m_D$ is comparable to the pion mass. If we define an effective mass $m_{eff}^2=\delta^2-m_\pi^2>0$, the OPE potential becomes
\begin{eqnarray}
  &V_{1\pi}&=-C\frac{(\epsilon\cdot p)(\epsilon'\cdot p)}{\bm{p}^2-m_{eff}^2-i0+},\label{eq:OPE1}\\
  &\text{Im }V_{1\pi}&=-\pi C(\epsilon\cdot p)(\epsilon'\cdot p)\delta(\bm{p}^2-m_{eff}^2).
\end{eqnarray}
We can find a pole on the real axis of the $p$-plane, which makes the potential non-Hermitian. And the potential in coordinate space oscillates after the Fourier transfromation \cite{Liu:2008fh}. 

We will encounter divergences if integrating along the real axis in Eq.~(\ref{eq:sch1}). To keep the analytical continuity, one should ensure that the pole does not cross the integral path during rotation, which demands $\theta>0$.

Compared with the instantaneous approximation, the OPE potential has a non-zero imaginary part proportional to $\delta(\bm{p}^2-m_{eff}^2)$. It results from the on-shell pion exchange, and will lead to a non-zero width of the bound state. 

According to the optical theorem, the imaginary part of the OPE potential is related to the three-body final state $DD\pi$ and the width arises from the decay mode $DD\pi$. Although the pole locates on the first Riemann sheet with respect to the $DD^*$ threshold, it may locate on the second Riemann sheet with respect to the $DD\pi$ threshold. In other words, it may be a $DD^*$ bound state and $DD\pi$ resonance.

\begin{figure}
  \centering
  \includegraphics[width=75mm]{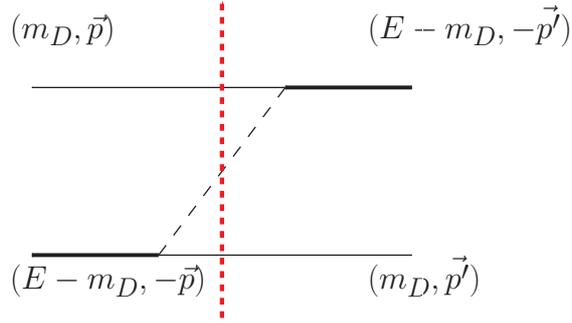}
  \caption{\label{fig:cutOPE} The one-pion-exchange diagram. The on-shell intermediate state contributes to the imaginary part of the potential. $E$ denotes the center-of-mass energy.}
\end{figure}

However, the above potential can not describe the behavior of the bound states near the $DD\pi$ threshold. In order to involve the three-body effect, we revise the OPE potential as
\begin{eqnarray}
  V_{1\pi}=C\frac{(\epsilon\cdot p)(\epsilon'\cdot p)}{(E+\delta)^2-\bm{p}^2-m_\pi^2+i0+},\label{eq:revOPE}
\end{eqnarray}
where $\delta$ is the mass splitting between the $D^*$ and $D$ mesons, and $E$ is the energy with respect to the $DD^*$ threshold, which is shifted compared with the notation in Fig.~\ref{fig:cutOPE}.  

The imaginary part of the revised OPE potential is coincident with the coupled-channel calculations. In the coupled-channel cases, the coupling to a lower channel will bring in an extra imaginary part to the effective potential of the higher channel.

\begin{figure}
  \centering
  \includegraphics[width=75mm]{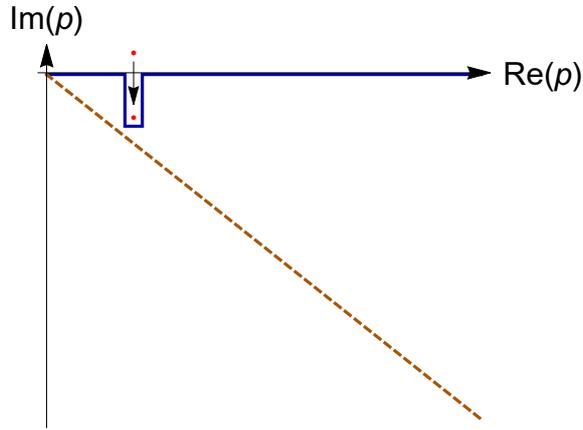}
  \caption{\label{fig:int}The integral path from $0$ to $\infty$ in the complex $p$-plane. The red point denotes the pole of the potential located at $p=\sqrt{(E+\delta)^2-m_\pi^2}$. When the pole passes across the positive real axis, we need to change the integral path to maintain the analytical continuity (blue solid curve). Instead, we can carry out a complex scaled integral (brown dashed curve) to deal with the pole.}
\end{figure}

When we put the intermediate $DD\pi$ on shell, we can derive Eq.~(\ref{eq:revOPE}) directly from Fig.~\ref{fig:cutOPE}. It looks kind of abnormal since the potential relies on the center-of-mass energy or binding energy. In fact, a unitary cut is introduced to the potential from $E=-\delta+m_\pi$ to $+\infty$. As shown in Fig.~\ref{fig:int}, the integral over $|\boldmath{p}|$ is usually applied from 0 to $+\infty$. However, when $E$ changes, the pole at $p=\sqrt{(E+\delta)^2-m_\pi^2}$ may cross the positive real axis. If we do not change the integral path, the discontinuity arises. This happens only when Re$(\sqrt{(E+\delta)^2-m_\pi^2})>0$. This is the origin why the potential has a branch point at the lower threshold $E=-\delta+m_\pi$. As indicated in Ref.~\cite{LHCb:2021auc}, the pole of the $T_{cc}^+$ state lies on the second Riemann sheet with respect to the $DD\pi$ threshold. To correctly put the potential onto the second Riemann sheet, we use the brown line as the integral path in Fig.~\ref{fig:int} instead of the blue one. In other words, the rotating angle $\theta$ in Eq.~(\ref{eq:sch2}) should be large enough for the $DD\pi$ resonances, which is different from the case of the $DD\pi$ bound states when a simple choice $\theta=0$ is permitted.

\section{Effective potential\label{sec:Veff}}
Compared with the Born approximation in the scattering, all two-particle-irreducible (2PI) diagrams sum up to the effective potential
\begin{eqnarray}
  V=-\frac{1}{4}\mathcal{M}_{\text{2PI}},\label{eq:effpotential}
\end{eqnarray}
where the factor $-\frac{1}{4}$ differs from the usual $-\prod_i\frac{1}{\sqrt{2M_i}}$ because of the normalization of the heavy meson fields.

Since ChEFT only works at the small momentum regions, we perform a Gaussian cutoff to regularize the effective potential $V$, which reads
\begin{eqnarray}
  \mathcal{F}(\bm{p},\bm{p}')=\exp\left[-(\bm{p}^2+\bm{p}'^2)/\Lambda^2)\right].\label{eq:regulator}
\end{eqnarray}

We demand $\mathcal{F}(\bm{p},\bm{p'})\rightarrow 0$ when $\bm{p},\bm{p}'\rightarrow\infty$ before and after the rotation in the complex plane to ensure that the Schr\"odinger equation can be solved numerically, which constrains the rotating angle $\theta<\pi/4$.

With the isospin symmetry, the potentials in the $D\bar{D}^*$ system can be obtained through a G-parity transformation. The result is summarized in Table~\ref{tab:Gtrans}, where the transferred momentum $p=p_1-p_4,\, q=p_1-p_3$. $p_1$ and $p_3$ denote the momenta of the initial and final $D(\bar{D})$ mesons, and $p_2$ and $p_4$ stand for the momenta of the initial and final $D^*(\bar{D}^*)$ mesons, respectively.

\begin{table}
  \caption{\label{tab:Gtrans}The pion-exchange potentials in the $DD^*$ and $D\bar{D}^*$ system are related by the G-parity transformation. "+" means the potential of the $D\bar{D}^*$ is the same as that of its partner in the $DD^*$ system with the same isospin, and "-" means an inverse in sign.}
  \centering
  \begin{tabular}{cccc}
      \hline
      &transfer momentum&G=+&G=-\\
      \hline
      OPE&p&+&-\\
      \hline
      TPE&q&+&+\\
      &p&-&+\\
      \hline
  \end{tabular}
\end{table}

To simplify the derivation, we only consider the LO potentials, namely the OPE potentials and the LO contact terms. The OPE potential has a generic form
\begin{eqnarray}
  V^i_{1\pi}=-\frac{g^2}{4f_\pi^2}\frac{(\epsilon\cdot p)(\epsilon'\cdot p)}{(E+\delta_i)^2-\bm{p}^2-m_\pi^2+i0+},\label{eq:revOPE}
\end{eqnarray}
where $E$ stands for the energy with respect to the lowest two-body threshold and $\delta_i$ stands for the mass difference in the $i$-th channel. We have the exchanged pion energy $p^0=E+m_{\text{th}}-m_{D,i}-m_{D,f}$ from the description in Fig.~\ref{fig:cutOPE}, where $m_{\text{th}}$ represents the mass of the threshold.  $m_{D,i}$ and $m_{D,j}$ stand for the masses of the initial and final $D$ mesons. For the one-eta-exchange diagrams, the subscript should be changed accordingly.

For the $DD^*$ system, we choose $m_{\text{th}}=m_{D^0}+m_{D^{*+}}$, and the OPE potentials are
\begin{eqnarray}
  &&V_{D^+D^{*0}\rightarrow D^+D^{*0}}=-2V_{1\pi^\pm},\nonumber\\
  &&\delta_{D^+D^{*0}\rightarrow D^+D^{*0}}=m_{D^0}+m_{D^{*+}}-2m_{D^+},\nonumber\\
  &&V_{D^+D^{*0}\rightarrow D^0D^{*+}}=V_{1\pi^0},\nonumber\\
  &&\delta_{D^+D^{*0}\rightarrow D^0D^{*+}}=m_{D^{*+}}-m_{D^+},\nonumber\\
  &&V_{D^0D^{*+}\rightarrow D^0D^{*+}}=-2V_{1\pi^\pm},\nonumber\\
  &&\delta_{D^0D^{*+}\rightarrow D^0D^{*+}}=m_{D^{*+}}-m_{D^0}.
\end{eqnarray}

For the $D\bar{D}^*$ system, we choose the states with the positive C-parity
\begin{eqnarray}
  &&|D^+D^{*-},C=+\rangle =\frac{1}{\sqrt{2}}\left(|D^+D^{*-}\rangle-|D^-D^{*+}\rangle\right),\nonumber\\
  &&|D^0\bar{D}^{*0},C=+\rangle =\frac{1}{\sqrt{2}}\left(|D^0\bar{D}^{*0}\rangle-|\bar{D}^0D^{*0}\rangle\right)\label{eq:c-parity}.
\end{eqnarray}

The corresponding threshold is $m_{\text{th}}=m_{D^0}+m_{D^{*0}}$ and the OPE potentials read
\begin{eqnarray}
  &&V_{[D^+D^{*-}]\rightarrow [D^+D^{*-}]}=-V_{1\pi^0},\nonumber\\
  &&\delta_{[D^+D^{*-}]\rightarrow [D^+D^{*-}]}=m_{D^0}+m_{D^{*0}}-2m_{D^+},\nonumber\\
  &&V_{[D^+D^{*-}]\rightarrow [D^0\bar{D}^{*0}]}=-2V_{1\pi^\pm},\nonumber\\
  &&\delta_{[D^+D^{*-}]\rightarrow [D^0\bar{D}^{*0}]}=m_{D^{*0}}-m_{D^+},\nonumber\\
  &&V_{[D^0\bar{D}^{*0}]\rightarrow [D^0\bar{D}^{*0}]}=-V_{1\pi^0},\nonumber\\
  &&\delta_{[D^0\bar{D}^{*0}]\rightarrow [D^0\bar{D}^{*0}]}=m_{D^{*0}}-m_{D^0},
\end{eqnarray}
where $[D^+D^{*-}]$ and $[D^0D^{*0}]$ are the shorthands of the $C=+$ states in Eq.~\ref{eq:c-parity}. The contribution of the one-eta-exchange potential is quite small and has little influence on the result.

Since we consider only the S-wave interactions, we perform the substitution $(\epsilon\cdot p)(\epsilon'\cdot p)\rightarrow \frac{1}{3}p^2(\epsilon\cdot\epsilon')$.

\begin{figure}
  \centering
  \includegraphics[width=28mm]{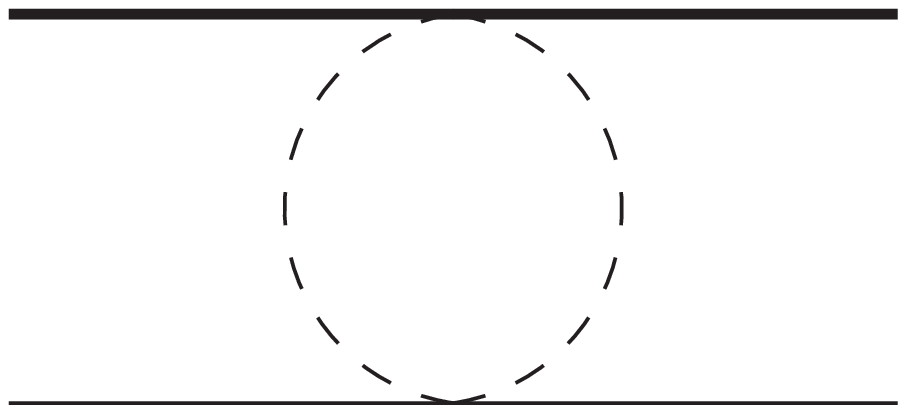}
  \includegraphics[width=28mm]{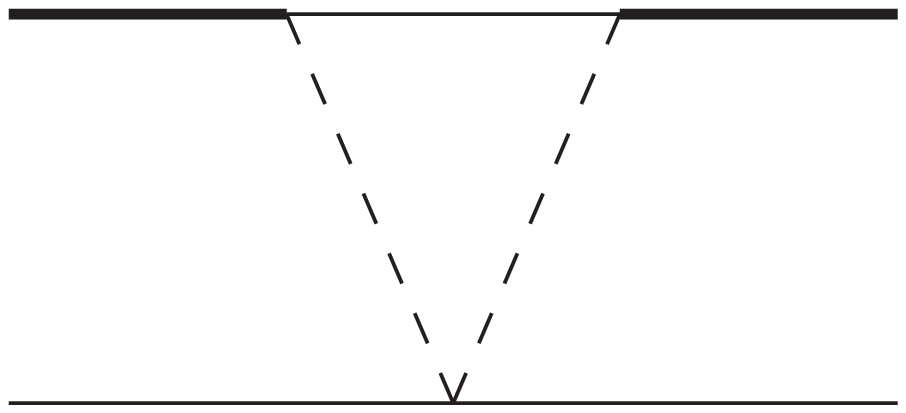}
  \includegraphics[width=28mm]{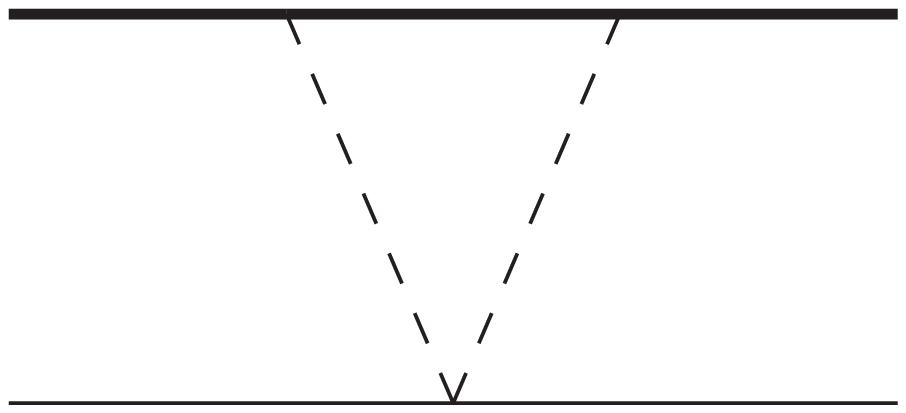}
  \includegraphics[width=28mm]{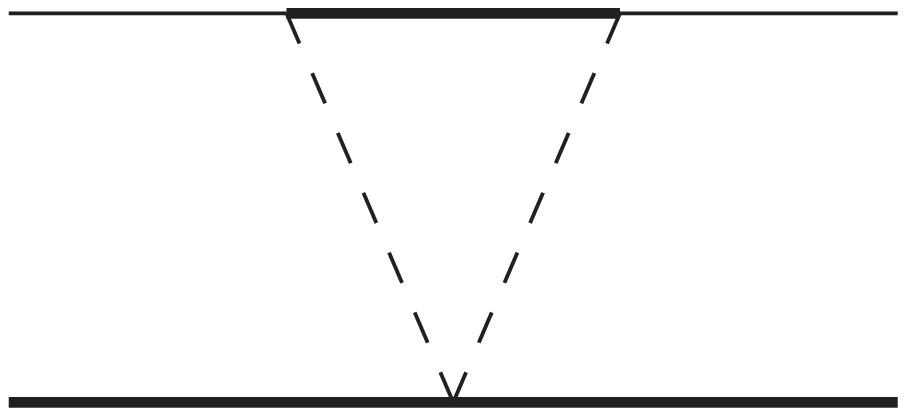}
  \includegraphics[width=28mm]{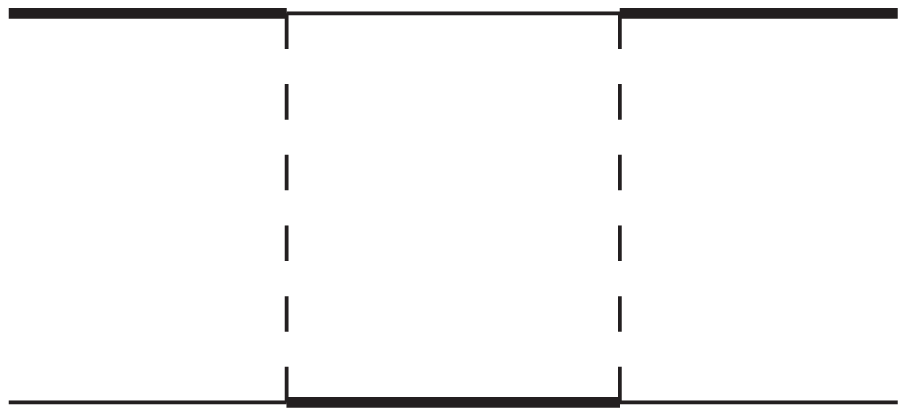}\vspace{5mm}
  \includegraphics[width=28mm]{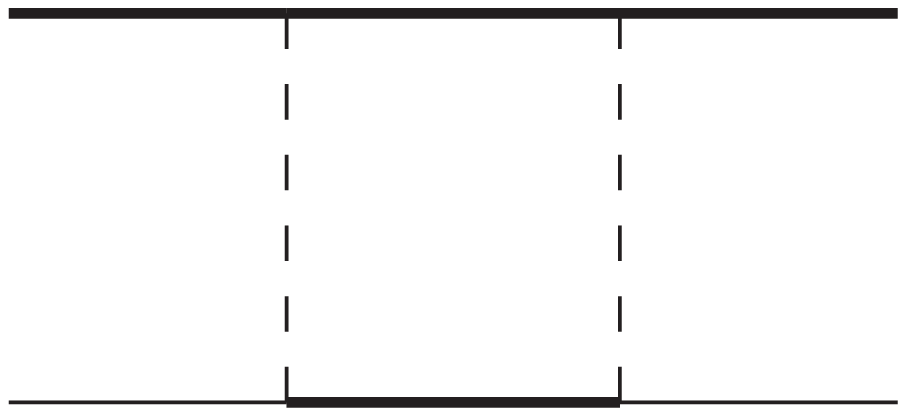}
  \includegraphics[width=28mm]{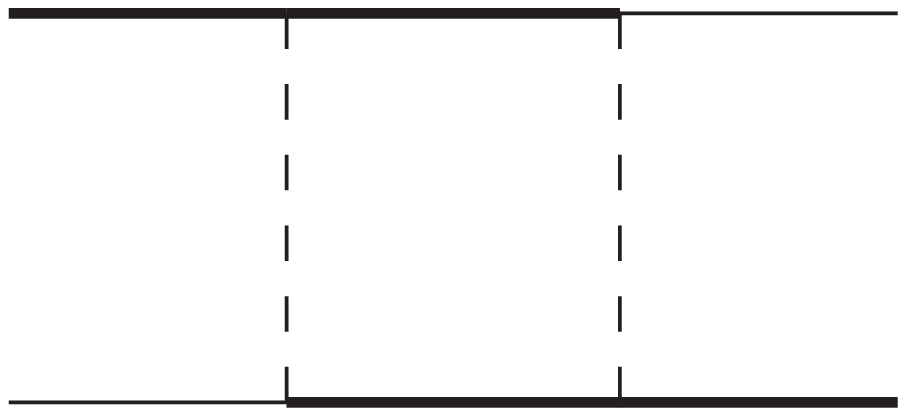}
  \includegraphics[width=28mm]{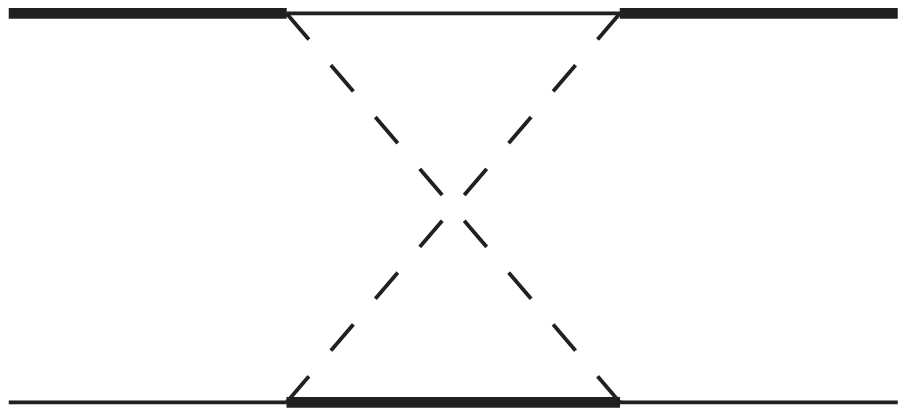}
  \includegraphics[width=28mm]{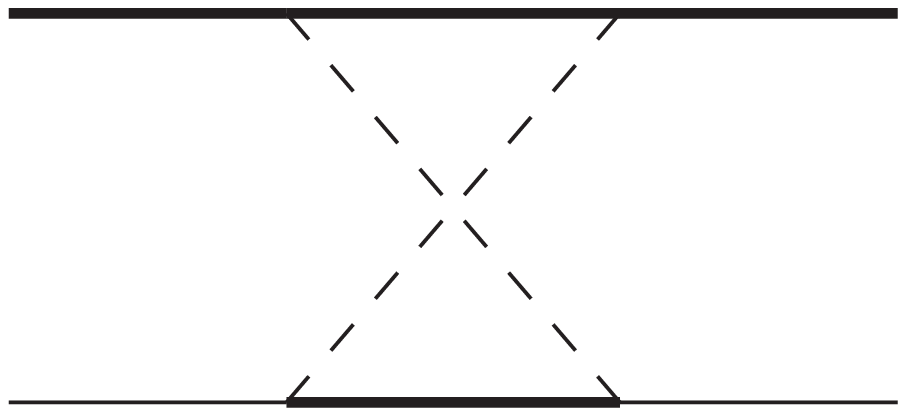}
  \includegraphics[width=28mm]{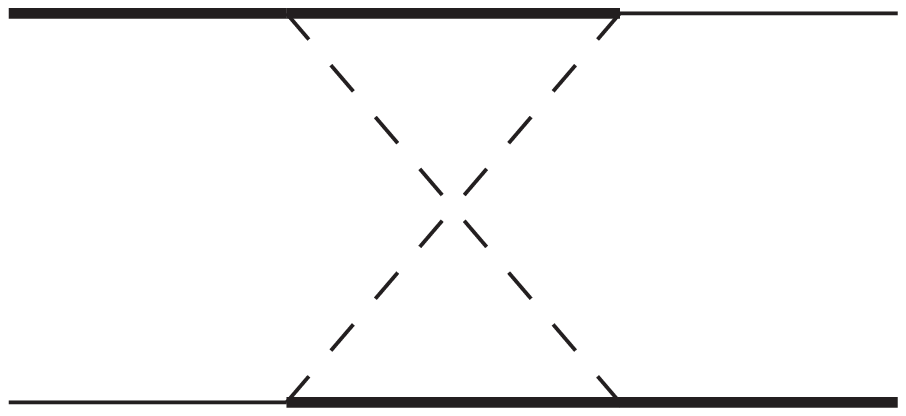}
  \caption{\label{fig:TPE}TPE diagrams. The thick, thin, dashed line represent $D^*$, $D$, $\pi$, respectively. The second, fifth, eighth diagrams have imaginary parts corresponding to on-shell intermediate states.}
\end{figure}

The TPE diagrams are shown in Fig.~\ref{fig:TPE}. The specific expressions can be found in Ref.~\cite{Xu:2017tsr,Wang:2018atz}. For planar box diagrams, we subtract the two-particle-reducible parts in line with the scheme in the appendix of Ref.~\cite{Wang:2019ato}. But we drop the fifth diagram, in which the intermediate $DD^*$ are on-shell. It has a large imaginary part which violates the unitarity condition.

\section{Numerical results and discussions\label{sec:numeric}}
\subsection{\label{sec:simple}Possible bound states with the $\mathcal{O}(p^1)$ $DD^*$ potential}
We first use the LO effective potentials only, i.e. the contact and OPE potentials, to show how the CSM works and clarify the $DD\pi$ three-body effect. Even this simplest potential can show the property of the $T_{cc}^+$. We choose the average value for the masses: $m_{D}=1.867$ GeV, $m_{D^*}=2.009$ GeV, $m_\pi=0.139$ GeV. The pion decay constant $f_\pi$ is chosen as 0.086 GeV and the $DD^*\pi$ coupling constant g is chosen as 0.65\cite{Xu:2017tsr}. The isospin breaking effect is not included.

\begin{figure}
  \includegraphics{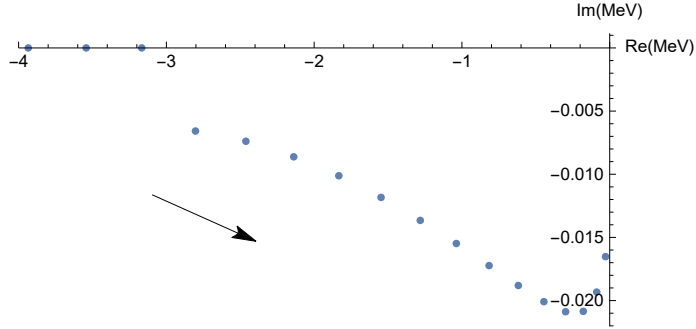}
  \centering
  \caption{\label{fig:polemovement}Bound states in the $I=0$ $S$-wave $DD^*$ system with the OPE and contact terms only. When we shift $C_s$ from -35 GeV$^{-2}$ to -19 GeV$^{-2}$, the pole moves in the positive real axis direction. With the parameters chosen, the $DD\pi$ threshold is located at -3 MeV. The cutoff $\Lambda$ is fixed to be 0.5 GeV.}
\end{figure}

\begin{table}
  \caption{\label{tab:estiLEC}Estimations of LECs with quark models (GeV$^{-2}$). }
  \centering
  \begin{tabular}{cccc}
    \hline
    &model I\footnote{Resonance saturation model. \cite{Epelbaum:2001fm,Xu:2017tsr}}&model II\footnote{Short-range quark-quark interactions via the fictitious scalar field and axial-vector field. \cite{Chen:2021cfl,Wang:2020dhf}}\\
    \hline
    $D_a$&-4.4&-2.0\\
    $D_b$&0&0\\
    $E_a$&-5.7&-5.9\\
    $E_b$&0&0\\
    \hline
    $\tilde{D}_a$&-6.7&-4.0\\
    $\tilde{D}_b$&0&0\\
    $\tilde{E}_a$&-5.7&-12.1\\
    $\tilde{E}_b$&0&0\\
    \hline
  \end{tabular}
  \end{table}
In some previous works, the LECs are estimated with the quark model such as the resonance saturation model, as shown in Table~\ref{tab:estiLEC}. Here we fix the LECs by fitting the experimental data. Fig.~\ref{fig:polemovement} shows the movement of the pole position with the varying LEC $C_s$, where $C_s=-2D_a+6E_a$. According to Table~\ref{tab:estiLEC}, we shift $C_s$ from -35 GeV$^{-2}$ to -19 GeV$^{-2}$. When $C_s$ increases, the potential becomes less attractive, and not strong enough to hold a stable bound state corresponding to a pole on the real axis. Then we can only find a pole in the second Riemann sheet with respect to the $DD\pi$ threshold. The imaginary part of the pole corresponds to the width decaying into the $DD\pi$ final state. However, we need to stress that it is still a (unstable) bound state of the $DD^*$ because they do lie on the physical sheet with respect to the $DD^*$ threshold. When fixing the real part of the pole to be -0.36 MeV, we have $C_s=-22.3\text{ GeV}^{-2}$ and the half width is determined to be $\Gamma/2=0.021\text{ MeV}$.

Generally speaking, the LECs depend on the cutoff $\Lambda$ in Eq~.(\ref{eq:regulator}) due to the renormalization. But the physical observables are independent of the cutoff chosen. If we change the cutoff and revise $C_s$ accordingly, it is interesting to see that the width is nearly unchanged, as shown in Table~\ref{tab:TccCutoff}. The width of the bound state is of the same order of magnitude as the width of the $D^*$ meson. As expected, it's  smaller than the $D^*$ width. In addition, no resonances are found under this set of parameters.

\begin{table}
\caption{\label{tab:TccCutoff}The dependence of the $I=0$ $DD^*$ contact interaction $C_s$ on the cutoff $\Lambda$ when the binding energy is fixed to the experimental value. Then the width is calculated using the determined $C_s$.}
\centering
\begin{tabular}{ccc}
\hline
$\Lambda$ (GeV)&$C_s$ (GeV$^{-2}$)&$\Gamma/2$ (MeV)\\
\hline
0.5 & -22.3 & 0.021\\
0.6 & -15.7 & 0.019\\
0.7 & -11.1 & 0.019\\
\hline
\end{tabular}
\end{table}

\subsection{Isospin breaking effects}
The isospin breaking effect needs to be included carefully because the characteristic energy scale $\delta-m_\pi$ is only several MeV, comparable with the mass splittings due to the isospin symmetry breaking. For example, the mass difference between the $D^+$ and $D^{*0}$ is 135.8 MeV. They can not exchange the on-shell $\pi^+$ under the heavy quark limit. Thus the potential in this channel has no singularity.

We perform the coupled-channel calculation involving the $D^+D^{*0}$ and $D^0D^{*+}$ with the parameters listed as follows: \cite{ParticleDataGroup:2020ssz}
\begin{eqnarray}
  &m_{\pi^\pm}=0.13957\text{ GeV},\qquad &m_{\pi^0}=0.13498\text{ GeV},\nonumber\\
  &m_{D^\pm} = 1.86966\text{ GeV},\qquad &m_{D^0} = 1.86484\text{ GeV},\label{eq:parameter}\\
  &m_{D^{*\pm}}= 2.01026\text{ GeV},\qquad &m_{D^{*0}}= 2.00685\text{ GeV}.\nonumber
\end{eqnarray}
The contact terms can be written as a matrix
\begin{eqnarray}
  V_{ct}=\begin{bmatrix}
    V_{D^+D^{*0}}&V_{D^+D^{*0}\rightarrow D^0D^{*+}}\\
    V_{D^0D^{*+}\rightarrow D^+D^{*0}}&V_{D^0D^{*+}}
  \end{bmatrix},
\end{eqnarray} 
which are related to 
\begin{eqnarray}
  V_{D^+D^{*0}}=V_{D^0D^{*+}}=\frac{1}{2}(V_{I=1}+V_{I=0}),\nonumber\\
  V_{D^+D^{*0}-D^0D^{*+}}=\frac{1}{2}(V_{I=1}-V_{I=0}),
\end{eqnarray}
where $V_{I=0}$ and $V_{I=1}$ stand for the LO LECs of the $I=0$ and $I=1$ channels, respectively.

Since there are two undetermined LECs, we can no longer give a prediction of the width. But we find the result is hardly sensitive to $V_{I=1}$, which implies the state is dominated by the $I=0$ channel. According to the quark model estimation, $V_{I=1}\approx20$ GeV$^{-2}$. Then $V_{I=0}$ is fixed to be -25.3 GeV$^{-2}$, which is in accordance with the simple calculation in \ref{sec:simple}. Now the width is nearly doubled and rises up to around 78 keV. 

\begin{table}
  \caption{\label{tab:TccIsobreak}Variation of the pole position $E=\delta m-i\Gamma/2$ (MeV) with respect to the $D^0D^{*+}$ threshold with the contact terms. It is sensitive to $V_{I=0}$ rather than $V_{I=1}$, which implies the pole is related to the $I=0$ state with a small isospin breaking effect. $V_{I=0}$ refers to $C_s$ in Table~\ref{tab:TccCutoff}.}
  \centering
  \begin{tabular}{clcccc}
    \hline
    &&\multicolumn{4}{c}{$V_{I=0}$ (GeV$^{-2}$)}\\
    &&-26&-25&-24&-23\\
    \hline
    \begin{tabular}{c}
      $V_{I=1}$\\(GeV$^{-2}$)
    \end{tabular}&\begin{tabular}{c}
      -10\\0\\10\\20\\30\\40
    \end{tabular}&\begin{tabular}{c}
      -0.537-0.039i\\-0.515-0.039i\\-0.502-0.039i\\-0.494-0.038i\\-0.488-0.038i\\-0.484-0.038i
    \end{tabular}&\begin{tabular}{c}
      -0.348-0.040i\\-0.325-0.040i\\-0.311-0.040i\\-0.303-0.039i\\-0.297-0.039i\\-0.292-0.039i
    \end{tabular}&\begin{tabular}{c}
      -0.190-0.041i\\-0.166-0.040i\\-0.153-0.040i\\-0.145-0.040i\\-0.139-0.039i\\-0.135-0.039i
    \end{tabular}&\begin{tabular}{c}
      -0.075-0.035i\\-0.054-0.033i\\-0.043-0.031i\\-0.037-0.030i\\-0.032-0.029i\\-0.029-0.029i
    \end{tabular}\\
    \hline
  \end{tabular}
  \end{table}

\subsection{Possible bound states in  the $D\bar{D}^*$ system}

Then we turn to the $D\bar{D}^*$ system, which is related to the $X(3872)$. With the exact isospin symmetry, the OPE potentials are exactly the same as their partners in the $DD^*$ system due to their positive G-parity (See Table~\ref{tab:Gtrans}). The only difference is the LECs in the contact Lagrangians. As we can see from Fig.~\ref{fig:polemovement}, the width of the $X(3872)$ is of the order of 10 keV assuming its decay is dominated by the $D\bar{D}\pi$ decay channel. 

However, the isospin breaking effect is significant for the $X(3872)$, since the mass difference between the $D^0\bar{D}^{*0}$ and $D^+D^{*-}$ is up to 8  MeV. The decay mode of the charged pions and $D$ mesons is kinetically forbidden. Thus the only possible on-shell pion is in the $D^0\bar{D}^{*0}$ channel.

\begin{figure}[htbp]
  \begin{minipage}{0.32\linewidth}
    \includegraphics[height=30mm]{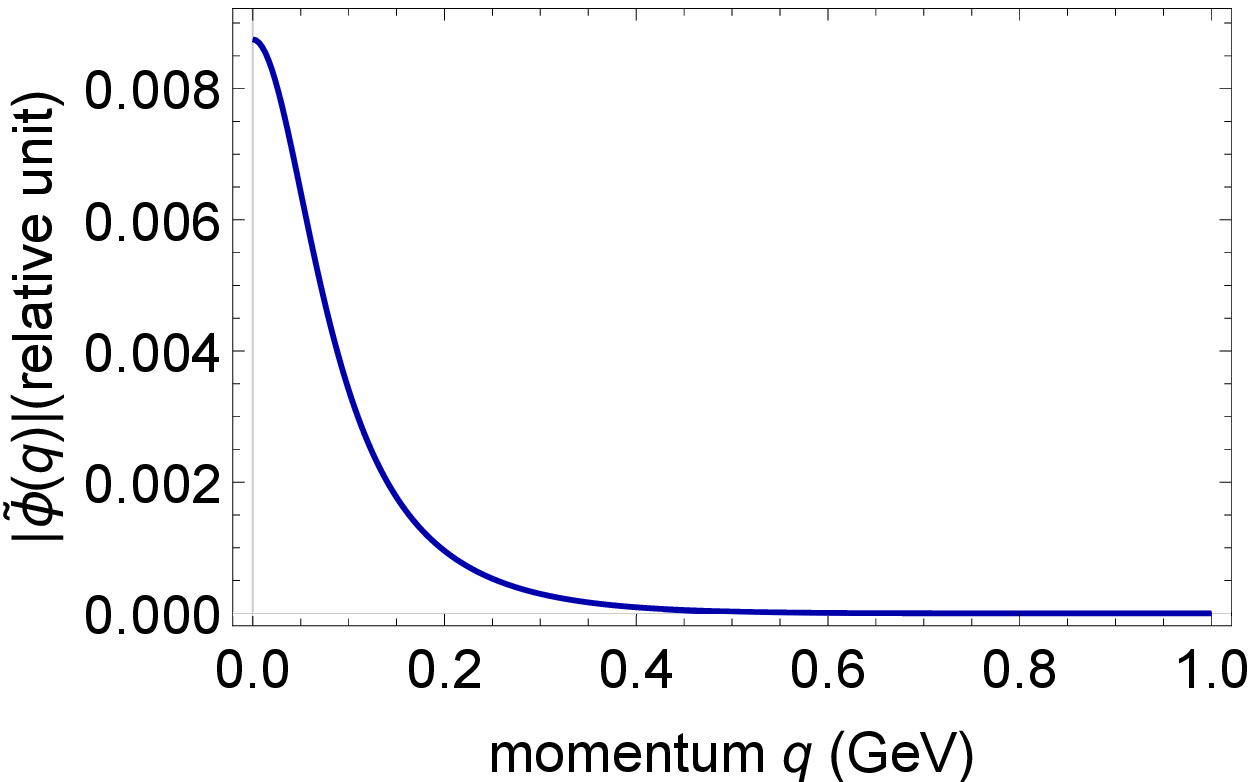}
  \end{minipage}
  \begin{minipage}{0.32\linewidth}
    \includegraphics[height=30mm]{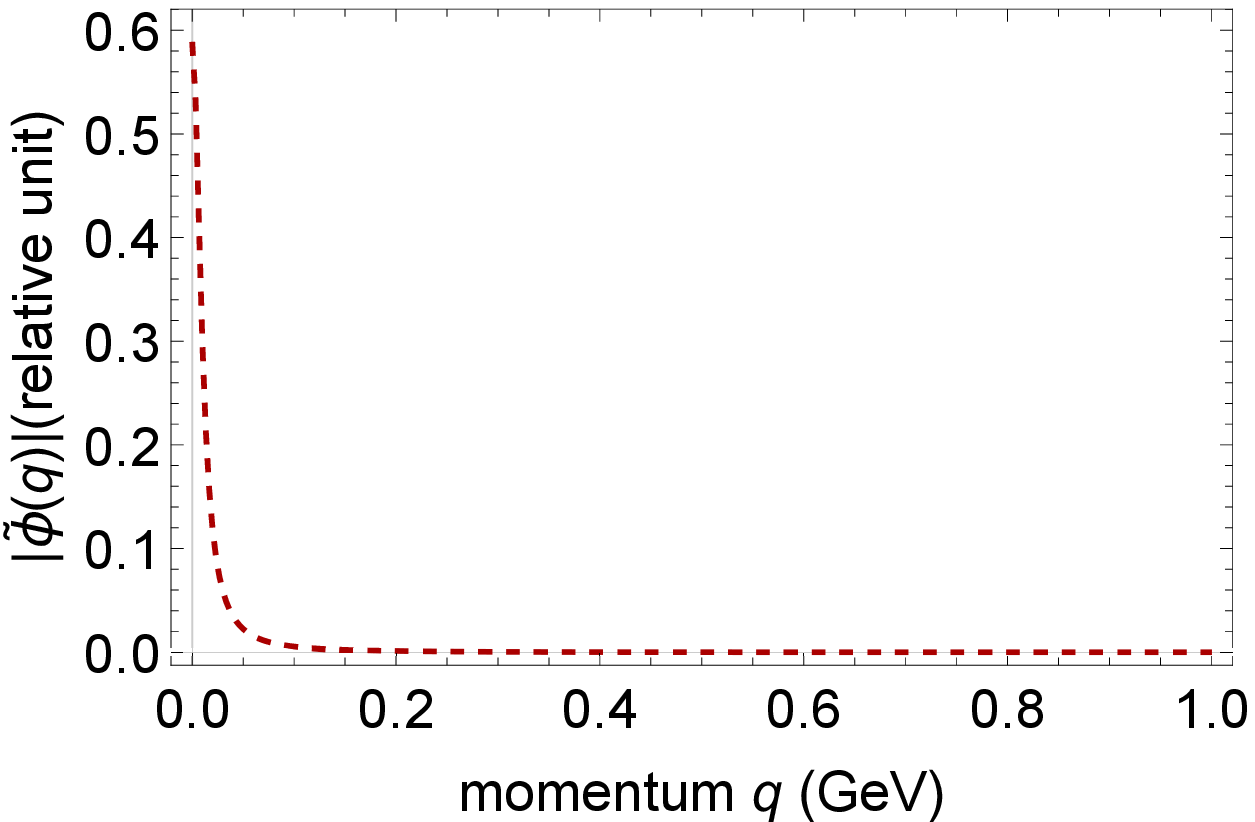}
  \end{minipage}
  \begin{minipage}{0.32\linewidth}
    \includegraphics[height=30mm]{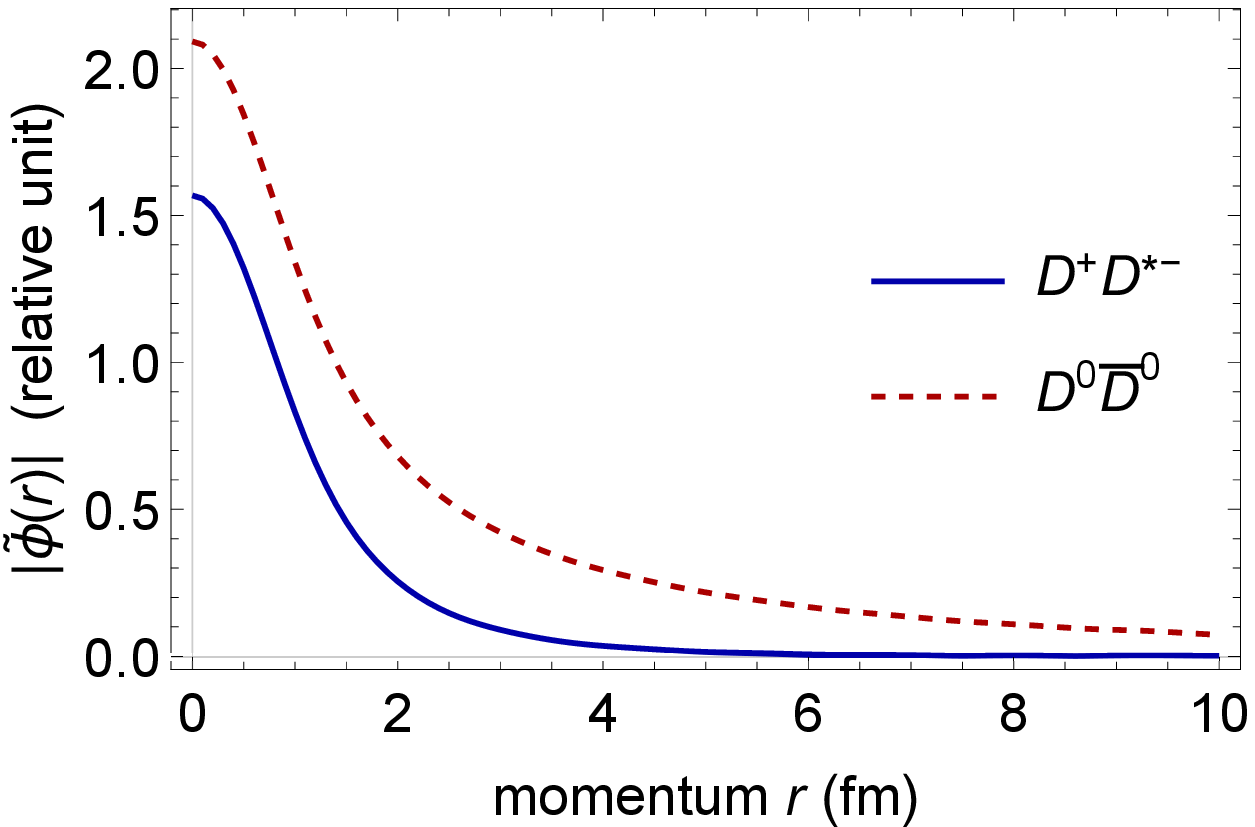}
  \end{minipage}
  \caption{\label{fig:wf3872}The complex scaled wavefunction $\tilde{\phi}_l(\bm{q})=\phi_l(\bm{q}e^{-i\theta})$ solved in Eq.~(\ref{eq:sch2}). The first two graphs stand for the $D^0\bar{D}^{*0}$ (red dashed curve), $D^\pm\bar{D}^{*\mp}$ (blue solid curve) channels, respectively. The third graph shows the complex scaled wavefunction $\tilde{\phi}_l(\bm{r})=\phi_l(\bm{r}e^{-i\theta})$ in coordinate space.}
\end{figure}

With the LECs in Table~\ref{tab:estiLEC}, we derive $V_{I=0}=2\tilde{D}_a+6\tilde{E}_a\approx -47$ GeV$^{-2}$ and $V_{I=1}=2\tilde{D}_a-2\tilde{E}_a\approx -2$ GeV$^{-2}$ from the resonance saturation model. The pole position is determined mainly by the $V_{I=0}$, but $V_{I=1}$ has a considerable contribution, implying a large isospin breaking compared with the $T_{cc}^+$. Fig.~\ref{fig:wf3872} shows the complex scaled wavefunction with $V_{I=0}=$-30 GeV$^{-2}$, $V_{I=1}=$5 GeV$^{-2}$ and $V_{I=0}=$-28 GeV$^{-2}$, $V_{I=1}=$5 GeV$^{-2}$. The pole lies at $E=(-0.114-0.017i)$ MeV. It is mainly a bound state of $D^0\bar{D}^{*0}$. The binding energy with respect to the $D^0\bar{D}^{*0}$ threshold is so small that the distribution of the wavefunction in momentum space is quite narrow in momentum space, and quite wide in coordinate space. The wave function seems comparable at small $\bm{r}$, but the wave function of $D^0\bar{D}^{*0}$ declines slowly as $r$ increases. The $D^0\bar{D}^{*0}$ channel takes up 94\% of the state. If the state is closer to the threshold, the isospin breaking will become larger.

When $V_{I=0}$ and $V_{I=1}$ vary, we find the width $\Gamma$ is always of the order of 10 keV, even smaller than the width of the $T_{cc}^+$, as shown in Table~\ref{tab:3872Isobreak}. So if we assume the $D^0D^0\pi^0$ and $D^0D^{*0}$ as the dominant decay channels, the width of the $X(3872)$ will be only of the order of 10 keV. Although this width looks too narrow, it is still within a standard deviation compared with the result of the pole search in Ref.~\cite{LHCb:2020xds}. The binding is so loose that the bound state may disappear if the potential is a little less attractive. The bound state pole can move above the $D^0\bar{D}^{*0}$ threshold with some specific sets of parameters, but we still recognize it as a (quasi-)bound state because it locates at the physical Riemann sheet.  

\begin{table}
  \caption{\label{tab:3872Isobreak}The variaion of the pole position $E=\delta m-i\Gamma/2$ (MeV) corresponding to the $D^0\bar{D}^{*0}$ with the contact terms. It is sensitive to both $V_{I=0}$ and $V_{I=1}$, which implies the isospin breaking effect is significant. The symbols are the same as in Table~\ref{tab:TccIsobreak}.}
  \centering
  \begin{tabular}{clccc}
    \hline
    &&\multicolumn{3}{c}{$V_{I=0}$ (GeV$^{-2}$)}\\
    &&-5&5&15\\
    \hline
    \begin{tabular}{c}
      $V_{I=1}$\\(GeV$^{-2}$)
    \end{tabular}&\begin{tabular}{c}
      -36\\-34\\-32\\-30\\-28
    \end{tabular}&\begin{tabular}{c}
      -1.591-0.014i\\-0.995-0.017i\\-0.517-0.020i\\-0.178-0.019i\\-0.008-0.012i
    \end{tabular}&\begin{tabular}{c}
      -1.498-0.014i\\-0.902-0.017i\\-0.431-0.019i\\-0.114-0.017i\\0.015-0.009i
    \end{tabular}&\begin{tabular}{c}
      -1.430-0.014i\\-0.834-0.017i\\-0.370-0.019i\\-0.074-0.016i\\-
    \end{tabular}\\
    \hline
  \end{tabular}
\end{table}

By analogy with the $N\bar{N}$ annihilation effects, we can introduce a complex $\tilde{C}_s$, according to the optical theorem, to take into account the inelastic channels, including the hidden-charm final states. This is based on the consideration that the annihilation processes of the quarks and antiquarks are usually related to the large momentum and short-ranged physics. Though we have little knowledge of how large the imaginary parts should be, we let them be of the same order of magnitude of their real parts, as we can see in the nucleon systems \cite{Kang:2013uia}. Due to the unitarity constraints, the imaginary part of the Feynman amplitudes satisfies $\text{Im } \mathcal{M}>0$. Then we demand $\text{Im } \tilde{C}_s<0$.

If we introduce the imaginary parts to the contact terms, the width of the $X(3872)$ will increase rapidly. For example, the pole will move to 0.10-0.39i MeV if we set $V_{I=0}=-30-3i$ GeV$^{-2}$ and $V_{I=1}=5$ GeV$^{-2}$. The paritial width of the $X(3872)$ decaying to the $D\bar{D}\pi$ final states is small due to the limited phase space. If the total width is much larger than the order of magnitude of 10 keV, it can not be explained by the three-body decay, and other decay channels including the hidden-charm decay modes are important.

\begin{figure}
  \centering
  \includegraphics[width=75mm]{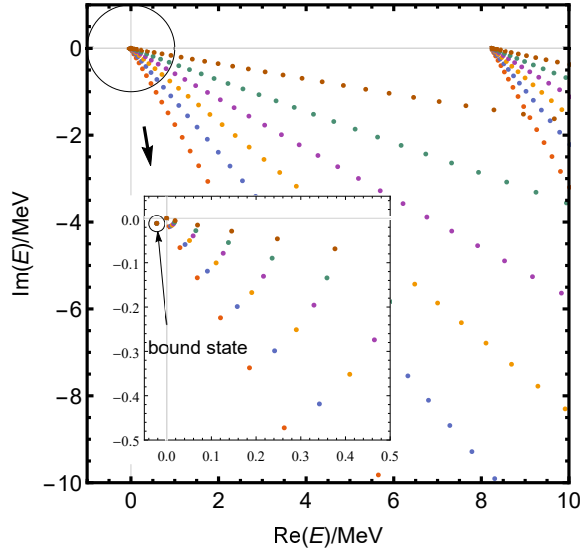}
  \caption{\label{fig:3872pole}Positions of the bound state corresponding to the $X(3872)$ found in the $D\bar{D}^*$ system. Dots with different colors represent the continuum states with the rotating angle $\theta=5^\circ, 10^\circ, 15^\circ, 20^\circ , 25^\circ, 30^\circ$. The dots on the left represent the $D^0\bar{D}^{*0}$ continuum and the dots on the right represent the $D^+\bar{D}^{*-}$ continuum. The lines always start from one of the thresholds. The unstable bound state lies below the line of the continum states and stays static when $\theta$ varies.}
\end{figure}

\subsection{Corrections from heavy meson kinetic energies}
In the framework of HMChEFT, the kinetic energies of heavy mesons are ignored. It is a reasonable approximation since the kinetic energies are small compared to the meson masses. However, the characteristic energy considered here is $\delta-m_\pi\approx 3\text{MeV}$, and the kinetic energy terms have a considerable contribution to the width of the state.\\

To evaluate the corrections from heavy meson kinetic energies, we revise Eq.~(\ref{eq:revOPE}) to its relativistic form. A relativistic kinetic energy term is introduced\footnote{Apart from the Lagrangians and the Feynman diagrams, Eq.~(\ref{eq:effpotential}) should be revised because of the normalization factor $-\prod_i\frac{1}{\sqrt{2M_i}}\rightarrow -\prod_i\frac{1}{\sqrt{2E_i}}$.}
\begin{eqnarray}
  \delta_i\longrightarrow m_D+m_{D^*}-\sqrt{\boldmath{p}_1^2+m_{D,i}^2}-\sqrt{\boldmath{p}_3^2+m_{D,f}^2},
\end{eqnarray}

As shown in Table~\ref{tab:kinetic}, we find the width drops by half when the kinetic energies included. Additionally, we find the relativistic form and the non-relativistic form of the kinetic energy show no difference.
\begin{table}
  \caption{\label{tab:kinetic}The width (unit: keV) of the pole found in the $DD^*$ and $D\bar{D}^{*}$ systems. The isospin conserved condition in the $D\bar{D}^{*}$ system is not included since the isospin breaking effect is large.}
  \centering
  \begin{tabular}{cccc}
    \hline
    &isospin conserving&isospin breaking&isospin breaking and kinetic energies\\
    \hline
    $DD^*$& 42& 78& 36\\
    $D\bar{D}^{*}$&-& 34& 15\\
    \hline
  \end{tabular}
\end{table}

\subsection{Possible bound states in  the $BB^*(\bar{B}^*)$ system}

Besides, we can use the above LECs to make predictions in the $BB^*(\bar{B}^*)$ system. Different from the $DD^*$ system, the OPE potential here has no singularity, and a generic Schr\"odinger equation without the three-body effect is enough if we only focus on the bound states. According to the heavy quark flavor symmetry, we presume the contact terms in the $BB^*(\bar{B}^*)$ system are the same as those in the $DD^*(\bar{D}^*)$ system. A $I=0$ and $J^P=1^+$ $T_{bb}^-$ pole is found at $E=-13.1$ MeV. An analog to the $X(3872)$ with $J^{PC}=1^{++}$ is found with its pole position $E=-14.6$ MeV corresponding to the $B\bar{B}^*$ threshold, which is mainly an isoscalar. The mass splitting from the isospin breaking effect in the $B\bar{B}^*$ system is much smaller and thus negligible.

\section{Two-pion-exchange potentials\label{sec:TPE}}
The imaginary part of the TPE potential naturally shows up in the loop integral. We include the TPE contributions and investigate its influence on the width of the $T_{cc}^+$ and the $X(3872)$. Its influence on the binding energy can be compensated by the adjustment of LECs. 

Since we calculate the TPE potentials under HMChEFT, namely all $1/M$ corrections are dropped, the kinetic energy corrections in OPE are not considered. 

\begin{figure}
  \includegraphics[height=50mm]{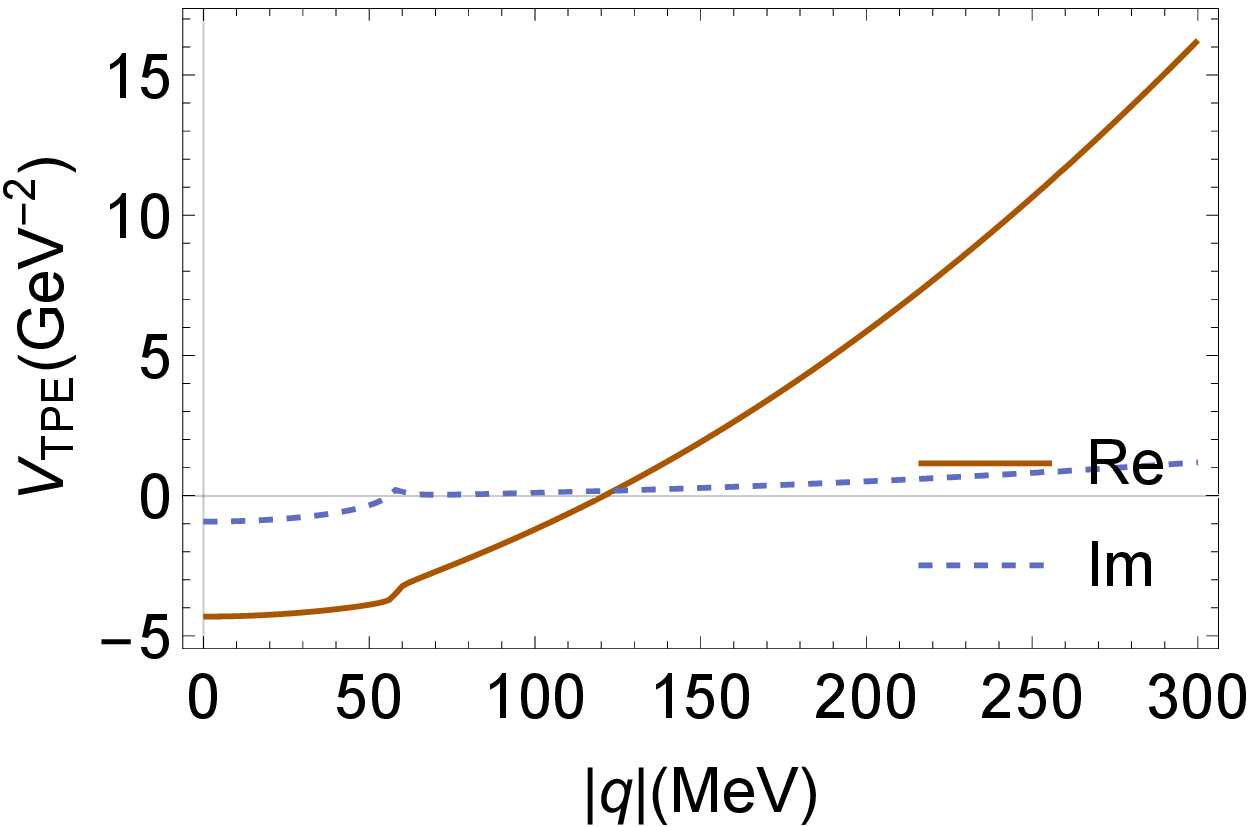}
  \includegraphics[height=50mm]{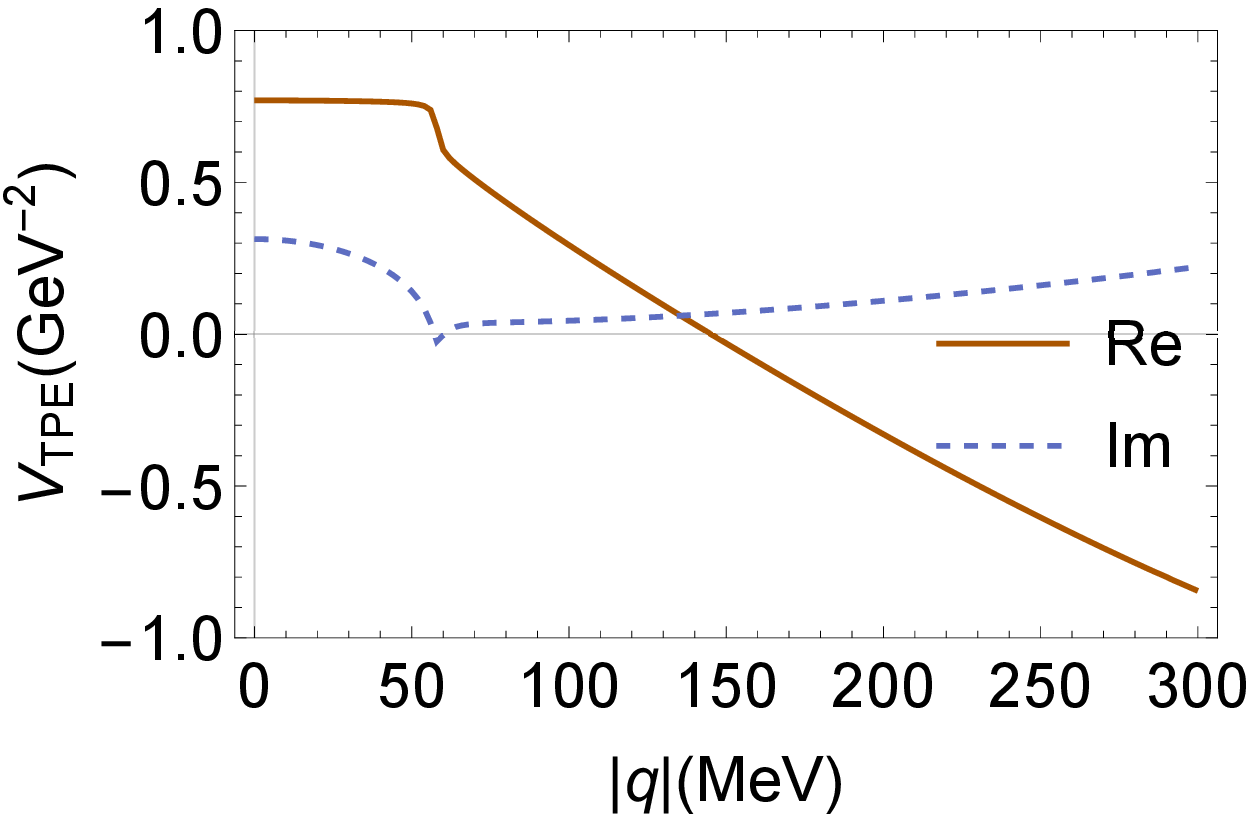}
  \caption{\label{fig:V2pi}The TPE potentials $V_{\text{TPE}}(\bm{q}e^{-i\theta})$ for $I=0$ (left) and $I=1$ (right) $DD^*$ systems when $\theta=1^\circ$. There is a pole near $|q|=60$ MeV corresponding to the on-shell pion if $\theta$ goes to zero.}
\end{figure}

\begin{figure}
  \centering
  \includegraphics[height=30mm]{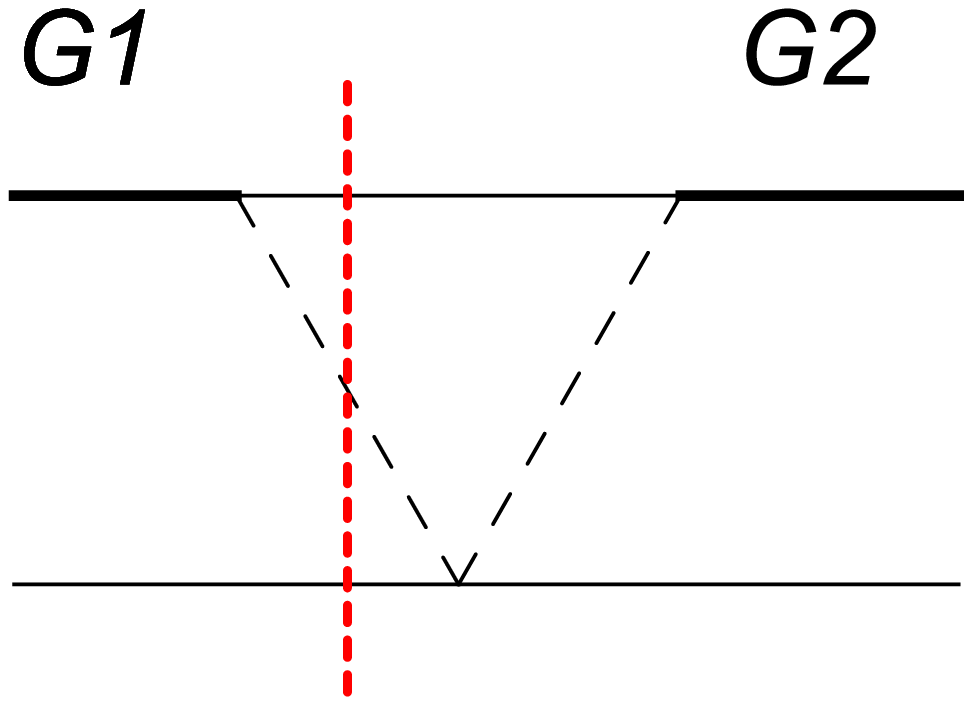}
  \includegraphics[height=30mm]{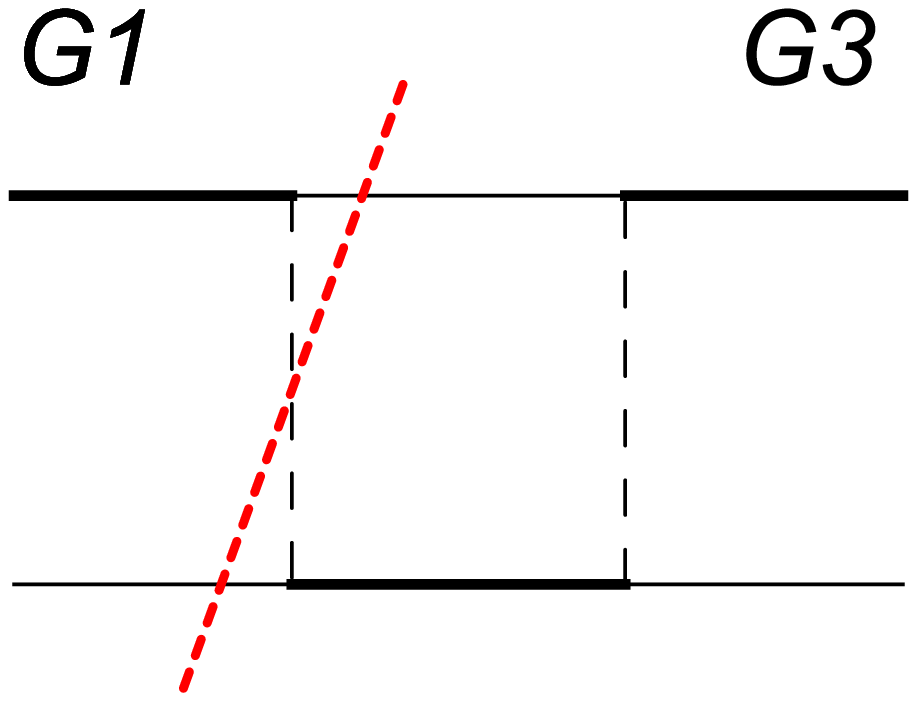}
  \caption{\label{fig:V2picut}Examples of on-shell intermediate states in TPE.}
\end{figure}
Fig.~\ref{fig:V2pi} shows the TPE potential versus the transferred momentum $\bm{q}$. In S-wave cases, it is equivalent to replace $\bm{p}$ with $\bm{q}$, so we simply plot the potential as a function of $\bm{q}$. There is a pole on the real axis at $|\bm{q}|=2\sqrt{\delta^2-m_\pi^2}$. We avoid the pole in the integral by rotating in the complex momentum plane. The imaginary part arises from the cut in Fig.~\ref{fig:V2picut}. However, for the subdiagrams $G_1$ and $G_2$, we only include $|G_1|^2$ and $G_1^*G_2$. $|G_2|^2$ is not included since it is of higher order. Then the unitarity is not guaranteed, which may result in a positive imaginary part of the potential. In Weinberg scheme, the 2PR part of the right diagram in Fig.~\ref{fig:V2picut} must be subtracted, but the previous subtraction scheme leads to a unphysically large imaginary part. So we exclude the diagram to keep unitarity. For similar reasons, one-loop corrections of OPE are not included.

\begin{figure}
  \centering
  \includegraphics[width=80mm]{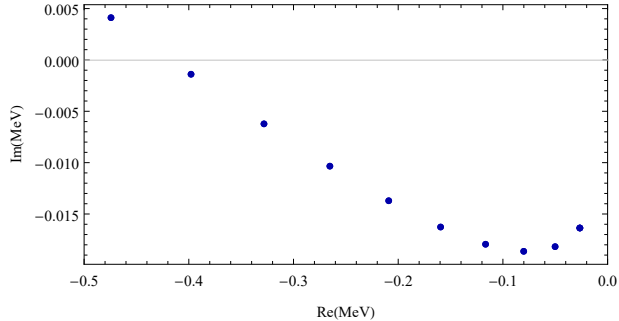}
  \caption{\label{fig:TPEpole}Bound states in the $I=0$ $S$-wave $DD^*$ system with the OPE, TPE and contact terms. When we shift $C_s$ from -41 GeV$^{-2}$ to -32 GeV$^{-2}$, the pole moves in the positive real axis direction.}
\end{figure}

Fig.~\ref{fig:TPEpole} shows the pole position of the $T_{cc}^+$ in $I=0$ channel as $C_s$ varies from -41 GeV$^{-2}$ to -32 GeV$^{-2}$. It seems the width declines and the pole can move to the upper half plane. But it is a result of the breaking of the unitarity, which is unphysical. In the framework of HMChEFT, one must be careful about the TPE when dealing with widths.

\section{Summary\label{sec:sum}}

In summary, we have calculated the pole position in the $DD^*$ system corresponding to the newly observed $T_{cc}^+$ state in the molecular state scenario with the LO OPE potentials and contact terms. The complex scaling method turns out to be a useful tool to involve the three-body effects into the Schr\"odinger equation. With a non-Hermitian effective potential, the Schr\"odinger equation has a solution on the first Riemann sheet, whose eigenenergy has a negative imaginary part, implying that the state decays.

With the energy of the exchanged pion retained, the OPE potential is dependent on the center-of-mass energy, which is similar to the coupled-channel analysis of the $DD\pi$ and $DD^*$ channels. This dependency introduces the unitary cut at the $DD\pi$ three-body threshold in the OPE potential, and influences the pole position. Finally, we find an unstable bound state with a finite width on the physical Riemann sheet with respect to the $D^0D^{*+}$ threshold.

For the calculations in the $DD^*$ system, we find only one pole in the first Riemann sheet with respect to the $DD^*$ threshold. It is a $DD^{*}$ bound state and $DD\pi$ resonance corresponding to the $T_{cc}^+$. We find neither $D^+D^{*0}$ bound state nor $D^0D^{*+}$ resonance. The width gradually goes to zero when the pole moves towards the $DD\pi$ threshold. But if we fix the real part of the pole position (binding energy), the imaginary part (width) shows little dependence of the cutoff or the contact term of the $I=1$ channel. After considering the isospin breaking effect, we find the width of the $T_{cc}^+$ to be about 80 keV.

As for the $X(3872)$, the isospin symmetry breaking effect is significant. Using estimations of the LECs from the quark model, we do find a pole corresponding to the $X(3872)$. Its width is much smaller than the fits either with the Breit-Wigner or Flatt\'e models, but within one standard deviation of the pole search. The phase space of the open-charm decay of the $X(3872)$ is limited and the partial decay width of the $D\bar{D}\pi$ mode is small. The electomagnetic decay need be considered in further study, together with the higher order chiral potentials.

\acknowledgments
This research is supported by the National Science Foundation of China under Grants No. 11975033, No. 12070131001 and No. 12147168. The authors thank G. J. Wang for helpful discussions.

\appendix
\section{Effects of $D^*$ width}

Given that $D^*$ has a non-negligible width, its width influences the width of the $T_{cc}^+$ state. It arises from the self-energy diagram of the $D^*$, which is different from Fig.~\ref{fig:cutOPE}. It comes out as a modification of the propagator of $D^*$ in Ref.~\cite{Du:2021zzh}. In other words, the width should be added to the kinetic terms of $D^*$ in the Lagrangian. Since the $D^*$ is off-shell, we include its width as a function of the center-of-mass energy rather than a constant. Then we modify the Schr\"odinger equation for unstable particles as
\begin{eqnarray}
  E\phi_{l}(p)=\left(\frac{p^2}{2m}-\frac{i}{2}\Gamma(E)\right)\phi_{l}(p)+\int\frac{p'^2dp'}{(2\pi)^3}V_{l,l'}(p,p')\phi_{l'}(p').\label{eq:sch-mod}
\end{eqnarray}

In principle, the width of $D^*$ can be calculated through the self-energy diagram, and the analytical form of its real part depends on regularization methods. Since $E$ is complex, the real part of $D^*$ self energy $\Sigma(E)$ may influence the width of the $T_{cc}^+$. But here we only consider the imaginary part of the self energy and assume the variaion of its real part is small. Then the $\Gamma(E)$ is proportional to the two-body phase space

\begin{eqnarray}
  &&\Gamma_{D^{*+}}(E)=\frac{g_A^2 m_{D^0} }{12\pi f_\pi^2 m_{D^{*+}}}k_{D^0\pi^+}^{3}+\frac{g_A^2 m_{D^+} }{24\pi f_\pi^2 m_{D^{*+}}}k_{D^+\pi^0}^{3},\nonumber\\
  &&\Gamma_{D^{*0}}(E)=\frac{g_A^2 m_{D^0} }{24\pi f_\pi^2 m_{D^{*0}}}k_{D^0\pi^0}^{3},
\end{eqnarray}
where $k_{D^0\pi^+}$, $k_{D^+\pi^0}$ and $k_{D^0\pi^0}$ are the momenta of final states. They are dependent on the total energy $E$. 

Since Eq.~\ref{eq:sch-mod} has not been discussed before, we leave the effect of $D^*$ width as uncertainties. In the leading order, $k_{ij}\propto(E-m_i-m_j)^{3/2}$. The influence of $\Gamma_{D^*}$ is estimated to be 4 keV.

\section{Estimations of LECs}
We use two models to estimate the LECs in the $DD^*(\bar{D}^*)$ system. In the single-channel case (Sec.~\ref{sec:simple}), there is only one independent LEC and it can be determined by the mass of $T_{cc}^+$. The fitted value is consistent with the estimated value. In coupled-channel cases, the number of LECs exceeds the number of observables. Therefore, we use the estimated LECs to give predictions. In principle, the LECs depend on the cutoff $\Lambda$. We fix the cutoff $\Lambda$ and let the LECs vary around the estimated values.

\subsection{Model I}
In line with Ref.~\cite{Xu:2017tsr}, the resonance saturation model is performed to determine the LECs. In this model, the contact interactions arise from meson changes including the $\rho$, $\omega$, $\sigma$ and other scalar or axial-vector mesons. The masses of mesons are relatively large, which results in a short-range interaction. We take the following substitution in propagators to generate contact interactions
\begin{eqnarray}
  q^2\longrightarrow 0.
\end{eqnarray}

In the one-boson-exchange model, the Lagrangian for the vector mesons reads
\begin{eqnarray}
  \mathcal{L}_{HHV}=-i\beta\langle Hv_\mu\rho^\mu\bar{H}\rangle+i\lambda\langle H\sigma_{\mu\nu}F^{\mu\nu}\bar{H}\rangle,
\end{eqnarray}
where $F^{\mu\nu}$ stands for the field-strength tensor $F^{\mu\nu}=\partial_\mu \rho_\nu-\partial_\nu \rho_\mu-[\rho_\mu,\rho_\nu] $. $\rho_\mu=\frac{ig_v}{\sqrt{2}}\hat{\rho}_\mu$ include $\rho$ and $\omega$ mesons under $U(2)$ symmetry
\begin{eqnarray}
  \hat{\rho}^\mu=\begin{pmatrix}
    \frac{\rho_0}{\sqrt{2}}+\frac{\omega}{\sqrt{2}} & \rho^+\\
    \rho^- &  -\frac{\rho_0}{\sqrt{2}}+\frac{\omega}{\sqrt{2}}
  \end{pmatrix}^\mu. 
\end{eqnarray}

The Lagrangians for the scalar and axial-vector mesons read
\begin{eqnarray}
  &&\mathcal{L}_{HHS}=g_{HHS}\langle HS\bar{H}\rangle,\\
  &&\mathcal{L}_{HHA_v}=g_{HHA_v}\langle H\gamma_\mu\gamma_5A_v^\mu\bar{H}\rangle,
\end{eqnarray}
where $S$ is the scalar field operator, and $A_v$ is the axial-vector field operator.

They contribute to the $D_a$ and $E_a$ terms according to the Lorentz structure and isospin. For the $DD^*$ system,
\begin{eqnarray}
  &D_a=-\frac{\beta^2g_v^2}{8m_\omega^2}+\frac{g_s^2}{2m_\sigma^2}+\frac{g_{s0}^2}{12m_{f_0}^2},\qquad &E_a=-\frac{\beta^2g_v^2}{8m_\rho^2}-\frac{g_{s0}^2}{4m_{a_0}^2},\\
  &D_b=\frac{g_{HHA}^2}{8m_{a1}^2},\qquad&E_b=\frac{g_{HHA}^2}{8m_{f1}^2}.
\end{eqnarray}
where $\beta g_v$, $g_s$, $g_{s0}$, $g_{HHA}$ are coupling constants for $\rho(\omega)$, $\sigma$ and $f_0(a_0)$, respectively. We take $\beta=0.9$, $g_v=5.8$, $g_s=0.76$, and $g_{s0}=\sqrt{3}g_{s}$ \cite{Li:2012ss,Liu:2008xz}. 

For the $D\bar{D}^*$ system, the $\omega/a_0/a_1$-exchange changes the sign,
\begin{eqnarray}
  &\tilde{D}_a=-\frac{\beta^2g_v^2}{8m_\omega^2}-\frac{g_s^2}{2m_\sigma^2}-\frac{g_{s0}^2}{12m_{f_0}^2},\qquad &\tilde{E}_a=-\frac{\beta^2g_v^2}{8m_\rho^2}+\frac{g_{s0}^2}{4m_{a_0}^2},\\
  &\tilde{D}_b=\frac{g_{HHA}^2}{8m_{a1}^2},\qquad&\tilde{E}_b=\frac{g_{HHA}^2}{8m_{f1}^2}.
\end{eqnarray}

Assumming LECs are saturated by the resonances below 800 MeV, we estimated the LECs including the contributions of the $\rho$, $\omega$ and $\sigma$ mesons,
\begin{eqnarray}
  &D_a\approx -4.4\text{ GeV}^{-2},\qquad &E_a\approx -5.7\text{ GeV}^{-2},\nonumber\\
  &\tilde{D}_a\approx -6.7\text{ GeV}^{-2},\qquad &\tilde{E}_a\approx -5.7\text{ GeV}^{-2}.
\end{eqnarray}

\subsection{Model II}
In Model II, we estimate LECs using quark-level Lagrangians. In Refs.~\cite{Chen:2021cfl,Wang:2020dhf}, the interactions between quarks are induced by the exchange of the fictitious scalar field $\mathcal{S}$ and axial-vector field $\mathcal{A}$,

\begin{eqnarray}
  \mathcal{L}=g_s\bar{q}\mathcal{S}q+g_a\bar{q}\gamma_\mu\gamma_5\mathcal{A}^\mu q.
\end{eqnarray}

The scalar field $\mathcal{S}$ and axial-vector field $\mathcal{A}$ are assumed to form a $SU(3)$ octet in flavor space. In $SU(2)$ case, they can be decomposed into isospin triplets and singlets,

\begin{eqnarray}
  &&\mathcal{S}=\mathcal{S}_3\tau^i+\frac{1}{\sqrt{3}}\mathcal{S}_1\tau^0,\nonumber\\
  &&\mathcal{A}^\mu=\mathcal{A}^\mu_3\tau^i+\frac{1}{\sqrt{3}}\mathcal{A}^\mu_1\tau^0,
\end{eqnarray}
where $\tau^0$ denotes the identity matrix. The $\sqrt{3}$ factor arises from the Gell-Mann matrix $\lambda_8$. 

Again the exchanged particles are assumed to be heavy. Then we obtain the contact interactions between the light quarks,

\begin{eqnarray}
    V_{qq}=c_s(1+3\tau_1\cdot\tau_2)+c_t(1+3\tau_1\cdot\tau_2)\sigma_1\cdot\sigma_2,\\
    V_{q\bar{q}}=\tilde{c}_s(1-3\tau_1\cdot\tau_2)+\tilde{c}_t(1-3\tau_1\cdot\tau_2)\sigma_1\cdot\sigma_2.
\end{eqnarray}
The coefficients $c_s$ and $c_t$ stand for the central potential and the spin-spin interaction, respectively. They are to be determined in the $NN(\bar{N})$ systems.

At the hadron level, the potential can be written in the form of the hadron spin and isospin, 
\begin{eqnarray}
  V_{DD^*}=c_s(1+3\tau_1\cdot\tau_2)
\end{eqnarray}
\begin{eqnarray}
    V_{D\bar{D}^*}=\tilde{c}_s(1-3\tau_1\cdot\tau_2)
\end{eqnarray}
\begin{eqnarray}
  V_{NN}=c_s(9+3\tau_1\cdot\tau_2)+c_t(1+\frac{25}{3}\tau_1\cdot\tau_2)\sigma_1\cdot\sigma_2
\end{eqnarray}
\begin{eqnarray}
    V_{N\bar{N}}=c_s(9-3\tau_1\cdot\tau_2)+c_t(1-\frac{25}{3}\tau_1\cdot\tau_2)\sigma_1\cdot\sigma_2
\end{eqnarray}

In Ref.~\cite{Chen:2021cfl}, the $qq$ interactions are determined by the experimental mass of the $P_c$ states,
\begin{eqnarray}
  c_s=3.9\text{ GeV}^{-2},\quad c_t=-0.95\text{ GeV}^{-2}.
\end{eqnarray}

In Ref.~\cite{Wang:2020dhf}, the $q\bar{q}$ interactions are determined by the scattering of $N\bar{N}$,
\begin{eqnarray}
  \tilde{c}_s=-8.1\text{ GeV}^{-2},\quad \tilde{c}_t=0.65\text{ GeV}^{-2}.
\end{eqnarray}

\bibliographystyle{JHEP}
\bibliography{Tcc}

\end{document}